Title: FePd$_3$ as a new material for thermally active artificial spin ice
Authors: Jasper Drisko[1], Stephen Daunheimer[2], John Cumings[2]
Affiliations:
[1]Department of Physics, University of Maryland, College Park, Maryland 20742, USA
[2]Department of Materials Science and Engineering, University of Maryland, College Park, Maryland 20742, USA

Abstract:

We introduce FePd$_3$ as a new material for studying thermally active artificial spin ice (ASI) systems and use it to investigate both the square and kagome ice geometries. We readily achieve perfect ground state ordering in the square lattice and demonstrate the highest yet degree of monopole charge-ordering in the kagome lattice. We find that smaller lattice constants in the kagome system generally produce larger domains of charge order. Monte Carlo simulations show excellent agreement with our data when a small amount of disorder is included in the simulation.

Main text:

Frustrated systems have emerged as an important topic of condensed matter physics, and geometric frustration is of particular prominence, where the frustration arises from an ordered structure rather than crystalline imperfections [1]. In such systems, an apparent degeneracy of ground states prevents long range order, often when detailed analysis of perturbations predict that an ordered state nevertheless should occur [2]. Despite decades of intense interest, frustrated systems still pose fundamental problems, with many unanswered questions, due in part to the tendency of these systems to inefficiently explore their configuration spaces and to lose ergodicity [3]. Monte Carlo simulations can address some of these issues, through the introduction of more complicated basic excitations [4,5], but questions about the specific conditions that will allow or restrict a system from entering into a predicted ground state often go unanswered in a general sense. An emerging approach seeks to address these issues by fabricating physical simulators of real materials [6–8], and artificial spin ices (ASI) comprise a highly productive example [9–14]. While it is now a topic of interest, the early historical work in ASI systems did not allow for thermalization of the material, and attempts to study the transition into the ground state had been carried out in these earlier studies only by AC demagnetization approaches [15,16], which do not access the long-range ordered states.

One seminal study found a ground state could be formed through thermalization occurring during the growth of the ASI magnetic nanostructure [13], but unfortunately this approach gives insufficient control over the process to address fundamental questions. More recently, a new experimental approach has been devised that seeks thermal activation in these physical simulators by heating close to the Curie temperature ($T_C$) of the constituent materials [17–25]. In this technique, the lithographically-patterned single-domain nanoscale bar magnets that comprise the structure may begin to exhibit spontaneous reversal, driven simultaneously by thermal effects and by the local magnetic environment. Work along these directions has shown long-range ordering in samples of the geometry known as square ice [17,19,23], and also has shown ordering of magnetic charges over comparatively short ranges in the kagome and shakti geometries [17,18,24,25]. However, the approach thus far has been limited by the choice of material and the nanoislands' geometries. Most studies have used permalloy ($Ni_{80}Fe_{20}$), which has a relatively high $T_C$, approximately 800 °C [26]. To accommodate such a high ferromagnetic ordering temperature, realizations utilize either narrow temperature ranges, which are limited at the upper end by degradation of the nanostructures [17,18], or very thin films (~3 nm) [19,20], which simultaneously reduce the onset temperature of thermally activated reversal and the desired magnetostatic interactions of the nanoscale bar magnets. Another technique is to use δ-doped Pd(Fe) monolayer stacks [21,22], which allow for a highly tunable $T_C$, but the films are still very thin (∼7 nm) and do not allow for strong magnetic coupling between macro-spins. A CoGd alloy has also been used to investigate the kagome lattice [24,25]. It has a relatively low $T_C$, around 200 °C, though the material is ferrimagnetic and the magnetization density is still significantly lower than that of the commonly-used permalloy.

In this work, we introduce a new, robust material for studying thermally activated artificial spin ices and demonstrate its value as an experimental tool. We readily reproduce previously reported perfect ground state ordering in the square geometry and present studies of the kagome lattice showing higher degrees of ordering in this fully frustrated system than previously reported. Additionally, we employ kinetic Monte Carlo modeling to interpret our results and illuminate the role of disorder in preventing perfect ordering of the kagome system. With the prospect that this disorder can be overcome, our system shows great promise for future studies by paving the way for even higher degrees of ordering and new explorations into more sensitive order parameters, closer to the true kagome ground state, and also more exotic geometries.

To fabricate our structures, we utilize a FePd alloy, close to the stoichiometry of $FePd_3$, which has not previously been used for such studies. An ideal material for thermal activation of ASI would exhibit simultaneously a high magnetic moment per atom and a relatively low $T_C$. The high moment allows for strong magnetostatic interactions between the macro-spin bar magnets. On the other hand, the lower $T_C$ – ideally not far above room temperature – allows accessing thermal activation while avoiding high temperatures where nanostructures are prone to degradation. According to binary alloy work [27], the $Fe_xPd_{1-x}$ system has a local maximum in the $T_C$, centered on the $FePd_3$ composition, so that fabrications of films within a wide range near this stoichiometric ratio, about 5%, still result in uniform properties and high experimental utility. The bulk material can have a moment density as high as 600 $kAm^{-1}$ [28] which is comparable to permalloy and other strongly magnetic materials. The $T_C$ of bulk $FePd_3$ is reported as ~260 °C [27], and the onset of spontaneous reversal in our nanostructures generally begins at about 110 °C, which leaves a large temperature window for experimental studies. Unlike previous studies, which have utilized thin magnetic layers with weak interactions [19–22] or thick layers and high temperatures [17,18], the studies we report here incorporate simultaneously low temperatures, thick layers, and strong magnetostatic interactions, which are all highly desirable features for a thermally active ASI.

We fabricate our samples on top of 100 nm thick, electron-transparent SiN membrane substrates, for studying the magnetic structures by Lorentz transmission electron microscopy (TEM). The bare substrates are first prepared by electron beam evaporating a 2 nm Ti adhesion layer and a 2 nm Pd wetting layer onto the SiN. A 23 nm $FePd_3$ film is then deposited by RF magnetron sputtering using a pure Fe target decorated with Pd pieces. After deposition, the films are annealed at 750 °C for 3 hours in UHV (base pressure $<6\times10^{-8}$ Torr) to help achieve favorable grain structure. Composition is characterized using wavelength dispersive x-ray spectroscopy on a JEOL JXA-8900 electron probe microanalyzer. Exact composition can vary slightly from run to run due to the nature of sputtering, but in a single deposition there is generally very uniform composition, only varying by 2-3% across the whole 3 inch deposition area. Figure 1(a) shows a Lorentz TEM image of a 23 nm thick $FePd_3$ film at room temperature. The textured bright and dark contrast in the image indicates the magnetization of the film is relatively high as the amount of contrast is proportional to the magnetization in the film [29]. Figure 1(b) shows that when the film is heated to 140 °C, the Lorentz contrast disappears, indicating the material is near its Curie temperature. The inset in Fig. 1(b) presents a high resolution TEM image showing the grain structure of the film. Grains are found to be 3-7 nm in size, which we note is favorable for thermal ASI since it will give rise to relatively low coercivity. We have tested our samples by heating them up to 510 °C and cannot detect any structural degradation at that temperature. This ASI system thus presents a much wider temperature window compared to permalloy [17] and allows for greater flexibility and ease in performing experiments. A comparison of the experimental temperature windows for the two materials is shown in Fig. 1(c).

To pattern the $FePd_3$ thin films into ASI geometries, we first spin a bilayer PMMA resist on top of the film and then perform electron beam lithography. A 12 nm Al etch mask is evaporated onto the sample and a liftoff step is performed. The pattern is then etched into the $FePd_3$ using Ar ion milling. Finally, a 5 nm Ti anti-charging layer is evaporated across the entire specimen. Images of the patterned films are shown in Figs. 2(a) and 2(c). The square ASI samples are disconnected ellipses 450 nm × 110 nm × 23 nm with a lattice constant of 500 nm. Kagome bar magnets are fabricated in a connected geometry, as we have previously reported [30,31], and vary in length from 500 nm to 300 nm. The width of the elements also varies, with smaller length magnets tending to be wider due to proximity effects in the lithography. The average widths are about 85 nm for the 500 nm long magnets up to 130 nm for the 300 nm length elements (see Fig. 6).

To observe thermal relaxation and investigate long-range ordering, samples are heated in Ar atmosphere above their $T_C$ and cooled at a rate of $1\,°C\,min^{-1}$ back down to room temperature. Before inserting samples into a transmission electron microscope for imaging, a degaussing procedure is run on the microscope's lens to remove any remnant field ($< 1$ gauss) at the sample that could influence the magnetic configuration. The samples are then imaged using Lorentz TEM [29–31]. Images are processed to automatically map the precise magnetic state of all macro-spins. Lorentz contrast images for the two geometries are shown in Figs. 2(b) and 2(d).

We observe the square geometry samples to readily find the ground state after annealing, often with perfect or nearly perfect ordering of type-I vertices. The largest perfect ground state domains we have observed are 30 μm × 30 μm, limited only by the size of the samples studied. An example of a perfectly ordered square sample is shown in Fig. 3. Under the same annealing conditions, the kagome geometry samples do not show the same transition to a long-range ordered state. Instead, we see domains of charge-ordering similar to Ref. [17], albeit with a greater degree of ordering, attributable to the more relaxed temperature requirements. Charge domain maps of kagome samples at different element lengths are shown in Fig. 4. It is apparent that shorter element lengths tend to result in larger domains of ordering. We also compute the charge correlation parameter C to quantify the ordering. This is found by assigning a 1(-1) to each pair of nearest neighbor vertices if they have the opposite (same) charge, adding up the contributions from all pairs and then dividing by the total number of nearest neighbors. We note that C=1 constitutes perfect ordering, C=0 is disordered, and C=-1 is anti-ordering. The best ordering we have observed is around C=0.6 in the 350 nm length samples, significantly higher than was previously reported in Ref. [17]. We also note that we do not observe any energetically unfavorable ±3 charges in our samples in contrast to other reports [17,18].

We have collected a number of data sets with different length elements, different crystals on one substrate, and different samples made from different depositions of FePd$_3$. When we average our correlation data at each length, a clear trend emerges: shorter lattice constants display greater degrees of charge-ordering. Reducing the element length increases the strength of the magnetostatic interactions between elements and the effective magnetic charges at the vertices, favoring lower energy configurations. The results of the full data set are shown in Fig. 5, and we discuss the deviation of the shortest elements in detail below.

To gain greater insight on charge-ordering and deviations from the predicted perfect order, we perform Monte Carlo simulations of our samples. Monte Carlo simulations have proven to be a successful approach in understanding complex behavior in other ASI systems [25,32–35]. We use a kinetic Monte Carlo technique [36] similar to other models of thermal ASI [19,20]. We calculate flipping rates as

$$\tau^{-1} = v_0 \exp(\frac{E_0 + \Delta E}{k_B T}) \tag{1}$$

where $E_0$ is the intrinsic energy barrier, $T$ is temperature, $v_0$ is a prefactor, and $\Delta E$ is the change in energy calculated from magnetic Coulomb interactions as in Refs. [14,37]. We are less concerned with the parameters $v_0$ and $T$ as these are used to evolve time in the model and affect every rate in the lattice in the same manner, whereas the present study focuses on how well the samples order at longer times. In our experimental samples, element width varies with length due to proximity effects during lithography. As such, we vary the element width with length in our simulations as $w = -0.2l + 190$. This is shown as the line in Fig. 6 and is representative of the widths of our real elements measured from high resolution TEM images.

With no disorder, the computational simulations produce perfect C=1 ordering every time, but we find that when we include a small amount of random disorder we are able to reproduce our experimental results quite well. There have been a number of studies focused on how disorder affects ground state formation in artificially frustrated systems [34,38–40]. In our samples, there is a disordered spread in the width of the magnets due to lithography artifacts, arising from slight local variations in electron dose due to shot noise. A 4 nm standard deviation of widths is representative of the real spread in element widths in our 500-350 nm length samples, from our microscopy observations (see Fig. 6). The 300 nm length samples are very difficult to fabricate and tend to have more edge roughness and a larger spread in element widths, closer to 7 nm. There were also fewer successful crystals made at this size. This explains the deviation of the 300 nm data point from the model.

Disorder in the widths of individual elements enters the simulations in two ways. First, the magnitude of the magnetic charge at each vertex is calculated from product of the magnetization and the cross sectional area, thus there are slight variations in the Coulomb energy interaction terms across the sample. Second, we estimate the intrinsic energy barrier $E_0$ for each element based on its shape anisotropy energy as in Refs. [20,23]. We calculate the demagnetizing factor (affected by the width of the specific element) and energy density (affected by demagnetizing factor) for each element and multiply by the individual element volume [41]. This leads to a range of $E_0$ values assigned to each spin, heavily affected by the actual width of the magnet. We run the simulations with the same number of elements (~5700) and same boundary conditions as our real samples. We use a magnetization of 200 kAm$^{-1}$ and evolve the simulation for an average of 200 flips per spin. We also repeat each point 75 times with different randomly distributed disorder and average over the results. When we assign each magnet a width randomly sampled from a Gaussian distribution with a standard deviation ($\sigma_w$) of 4 nm, we find very good agreement with our experimental data. By increasing the standard deviation of the distribution to 5nm, we are able to reproduce the 300 nm data point and also demonstrate the same upward trend in the correlation as a function of decreasing element length. The results of these simulations are presented in Fig. 5.

We also perform similar kinetic Monte Carlo simulations for the square geometry. Here, instead of the Coulomb energy we include magnetic dipolar interactions as in Ref. [19]. We find that with the same choice of parameters and same amount of width disorder (4 nm) as the kagome simulations, we still get very good long-range ordering and large domains of ground state type-I vertices in the square lattice, similar to the experiments shown in Fig. 3. This shows that the critical slowing down of spin dynamics in the kagome geometry [42] may be more susceptible to disorder, further preventing perfect long-range ordering in this system.

In future work, we hope to achieve even further ordering than presented here. A two stage ordering process is predicted for the kagome geometry, characterized by two plateaus in the entropy of the system and two corresponding phase transitions [5]. The first plateau is the magnetic charge-ordered state, followed by a further reduction of entropy into a spin-ordered state. This state could be achieved by reducing the disorder in our samples. Careful studies of resist thickness, dose, and accelerating voltage in the lithography step will help reduce the spread in widths of individual elements. Also Ar ion milling is an inherently disordered technique and new ways to transfer the ASI pattern into the $FePd_3$ films could be explored. In light of the success in easily achieving the square geometry ground state and our very well charge-ordered kagome lattices, we believe $FePd_3$ creates a promising pathway for reaching the first plateau with perfect charge-ordering, paving the way for future studies probing fundamental kinetics of the second phase transition. The experimental observation of the nature of this phase transition is of great interest and will shed light on analogous phase transitions in real frustrated materials like the Pyrochlore spin ices.

This work was supported by NSF CAREER Grant No. DMR-1056974. We also acknowledge the support of the Maryland NanoCenter and its NispLab and FabLab.

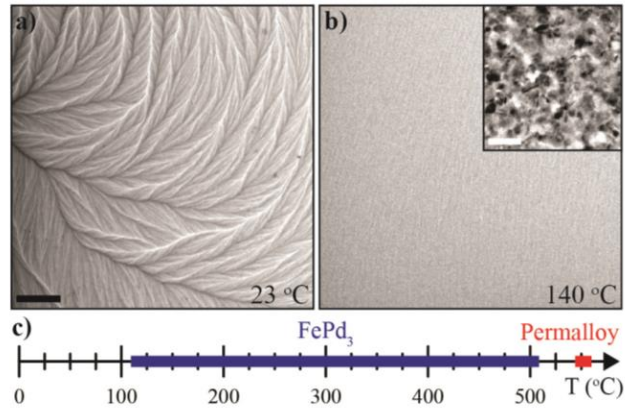

FIG. 1. Lorentz contrast transmission electron microscope (TEM) images of a 23 nm FePd$_3$ thin film at (a) room temperature and (b) 140 °C. The textured contrast in (a) indicates the film has a high magnetization, while the absence of contrast in (b) signifies the material is near its Curie point (scale bar: 20 μm). Inset: High resolution TEM image showing the grain structure of the film. Grains are 3-7 nm in diameter (scale bar: 25nm). (c) Experimental temperature window comparing FePd$_3$ and 25 nm thick permalloy. Lower bound is the temperature where magnets begin to flip due to thermal activation, higher bound is limited by nanostructure degradation. Permalloy data reported in Ref. [17].

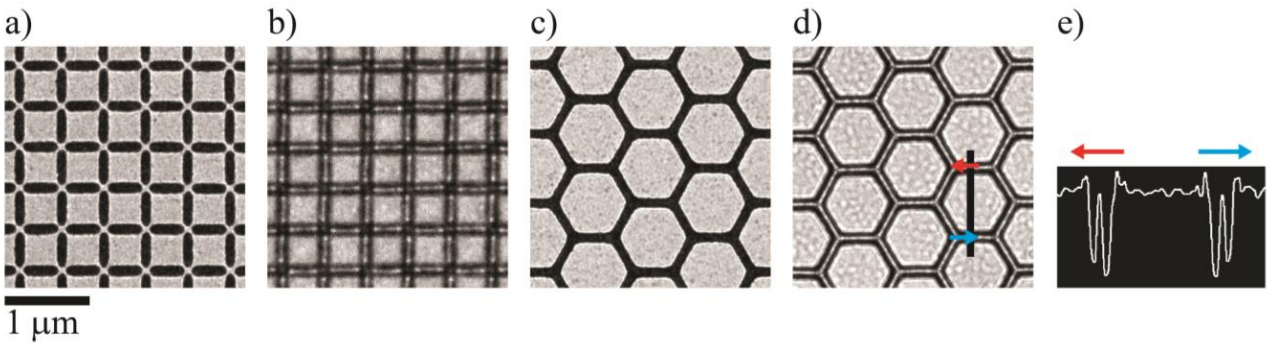

FIG. 2. (a) In-focus TEM image of a square geometry sample. (b) Lorentz contrast TEM image of the same square geometry sample, showing ground-state order. (c) In-focus TEM image of a kagome geometry sample. (d) Lorentz contrast TEM image of the same kagome geometry sample, showing only ice-rule ordering. (e) Intensity profile along the black line in (d) with corresponding arrows showing how the asymmetry in contrast across each element is used to determine the direction of the magnetic moment.

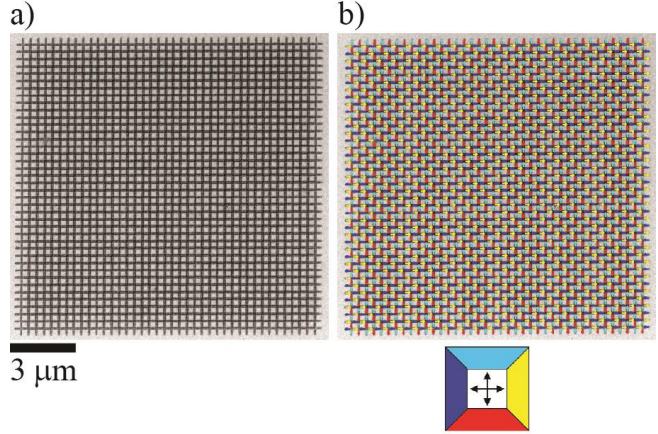

FIG. 3. (a) Lorentz contrast TEM image and (b) corresponding spin map of a square geometry sample that was heated above its $T_C$ and cooled back to room temperature showing perfect ground state ordering of type-I vertices. Such behavior is robust among the samples studied herein.

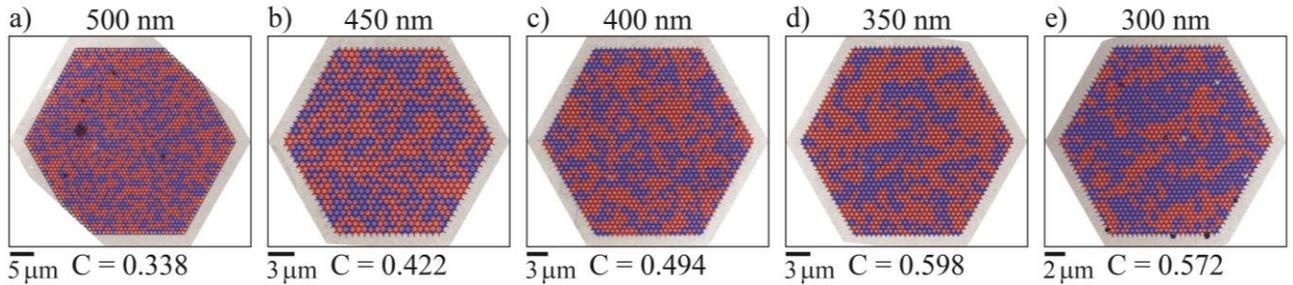

FIG. 4. Charge domain maps overlaid on top of Lorentz TEM images of kagome geometry samples of varying element length from 500 nm to 300 nm. Shorter element lengths tend to lead to greater degrees of charge-ordering and larger charge domains. We note that a magnetically saturated state would also be highly charge-ordering, however our samples are necessarily demagnetized, with the net magnetization $|M_x| < 0.2$ for all data. Also see Fig. 1 for a close up of a Lorentz image without charge domains.

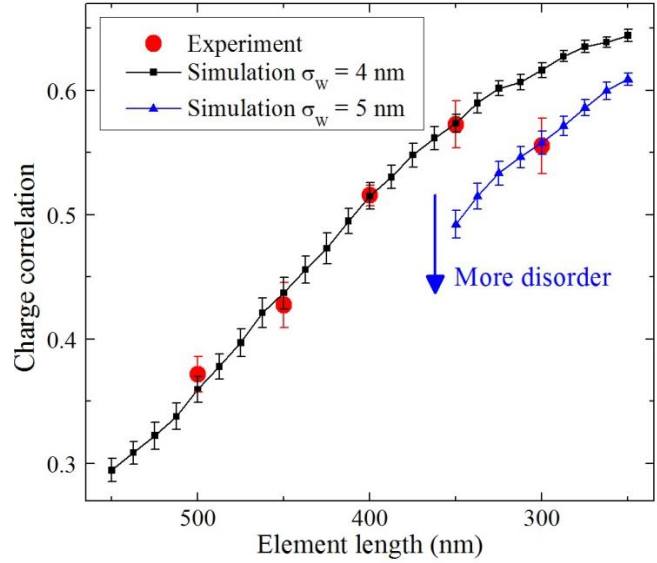

FIG. 5. Charge correlation as a function of element length in the kagome geometry for our experimental results and kinetic Monte Carlo simulations including two levels of width disorder. Experimental error bars are three times the uncertainty in the average, simulation error bars are two times the uncertainty in the average.

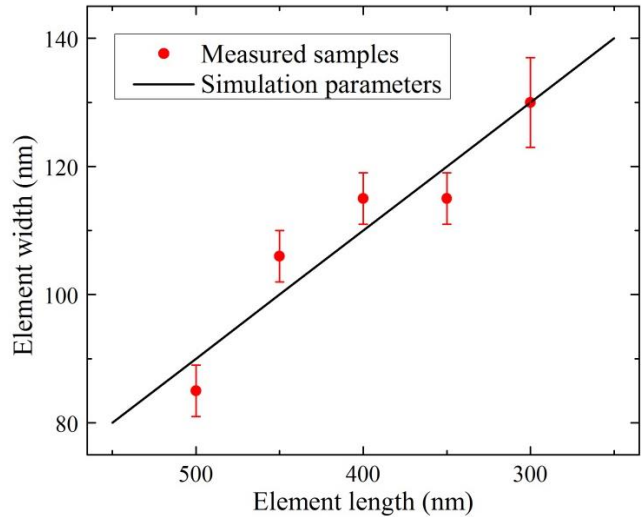

FIG. 6. Element width vs. element length measured from TEM images of our samples and a line showing correlated length and width input parameters in our simulations. Error bars are one standard deviation of measured widths. Note the 300 nm length elements have a larger spread in widths than the other lengths.